# Non-Equilibrium First-Order Exciton Mott Transition at Monolayer Lateral Heterojunctions Visualized by Ultrafast Microscopy

Long Yuan[1#], Biyuan Zheng[2#], Qiuchen Zhao[1#], Roman Kempt[3], Thomas Brumme[3], Agnieszka Beata Kuc[4*], Chao Ma[2], Shibin Deng[1], Anlian Pan[2*] and Libai Huang[1*]

[1]Department of Chemistry, Purdue University, West Lafayette, IN, USA

[2]Key Laboratory for Micro-Nano Physics and Technology of Hunan Province, State Key Laboratory of Chemo/Biosensing and Chemometrics, College of Materials Science and Engineering, Hunan University, Changsha, People's Republic of China

[3]Technische Universitat Dresden, 01062 Dresden, Germany

[4]Helmholtz-Zentrum Dresden-Rossendorf, Abteilung Ressourcenökologie, Forschungsstelle Leipzig, Permoserstr. 15, Leipzig, Germany

[#]These authors contribute equally.
*Corresponding authors: libai-huang@purdue.edu; anlian.pan@hnu.edu.cn; a.kuc@hzdr.de

## Abstract

Atomically precise lateral heterojunctions based on transition metal dichalcogenides provide a new platform for one-dimensional (1D) excitons. Here, we present the transport of different phases of the 1D interfacial excitons in a type II WSe$_2$-WS$_{1.16}$Se$_{0.84}$ lateral heterostructure using ultrafast microscopy with ~ 200 fs temporal resolution and nanoscale spatial resolution. These measurements revealed a highly non-equilibrium first-order exciton Mott transition at a density of ~ $5 \times 10^{12}$ cm$^{-2}$ at room temperature. The rapid expansion of dense electron-hole (e-h) plasma with a velocity up to $3.2 \times 10^6$ cm s$^{-1}$ was directly visualized. Below the Mott transition, repulsive exciton-exciton interactions led to enhanced exciton transport along the interface. These results highlight that atomically thin lateral heterojunctions can be designed as novel "highways" for excitons and dense e-h plasma for high-speed electronic applications.

# Main

Exciton Mott transition, at which a dilute exciton gas converts to a dense electron-hole (e-h) plasma, is a long-standing subject in the many-body physics of photoexcited semiconductors[1-4]. Two-dimensional (2D) transition metal dichalcogenide (TMDC) layers and their heterostructures, that support strongly bound excitons, have emerged as an exciting material platform for engineering exciton phase diagrams at room temperatures[5-13]. Recently, major advances have been made in the growth of a relatively new type of heterostructures, namely lateral heterostructures where two TMDC monolayers are covalently stitched in the 2D plane[14-21]. In these heterostructures, one-dimensional (1D) bonded interfaces are formed instead of the more extensively studied 2D van der Waals interfaces of vertically stacked heterostructures. Theoretical investigations have predicted spatially separated charge transfer (CT) excitons at the lateral interfaces with binding energy on the order of 200 meV as the ground state of the heterostructures[22,23].

The interfacial excitons at these lateral interfaces have unique properties that are promising for realizing 1D Mott transition at room temperature[23,24]. First, long lifetime is expected due to the spatial indirect nature of the interfacial CT excitons. Second, these interfaces are effectively an energy sink, because the interfacial excitons have lower energy than the intralayer counterparts, which can be leveraged to sustain a high carrier density. These unique properties could overcome some of the difficulties encountered in realizing Mott transition in other 1D systems, such as single-walled carbon nanotubes, where it is challenging to achieve high carrier density due to efficient exciton annihilation[25]. Thus far, Mott exciton transition in 1D has almost exclusively been investigated in semiconducting quantum wires at cryogenic temperatures[26].

However, probing exciton phase transitions in spatial regions of only a few nanometers presents significant technological challenges. In addition, Mott transition is inherently a non-equilibrium phase transition due to the ultrafast dynamics of the excitons and e-h plasma. To address these challenges, we employed here ultrafast pump-probe microscopy[27-29] with a temporal resolution of ~ 200 fs and a spatial precision of ~ 50 nm to image the transport of excitons and dense e-h plasma at the atomically sharp interface of $WSe_2$-$WS_{2(1-x)}Se_{2x}$ lateral heterostructures. The electronic structure of the interfacial excitons was elucidated by density functional theory (DFT) calculations. Ultrafast microscopy measurements revealed a non-equilibrium first-order Mott transition, showing an abrupt increase in the carrier diffusivity from an insulating exciton phase to a metallic e-h plasma phase. The collective excitation of plasma propagated with a velocity as high as $3.2 \times 10^6$ cm s$^{-1}$ over micrometer distances. Below the Mott transition, strong exciton-exciton repulsion led to an enhanced exciton migration along the interface.

**Structure of $WSe_2$-$WS_{2(1-x)}Se_{2x}$ lateral heterostructures.** To study exciton phase transition at 1D TMDC interfaces, we performed measurements on $WSe_2$-$WS_{2(1-x)}Se_{2x}$ lateral heterostructures grown by a modified chemical vapor deposition (CVD) method that provides a precise spatial modulation in chemical component and atomically sharp interfaces (Methods and Supplementary Fig. 1)[19]. $WSe_2$/$WS_{2(1-x)}Se_{2x}$ lateral heterostructures, with $x$ of 0.42 ($WSe_2$-$WS_{1.16}Se_{0.84}$, Supplementary Fig. 2), were selected for their improved stability over $WSe_2$/$WS_2$ interface under intense laser excitation. A sharp interface with low defect density is crucial for achieving energy confinement in 1D[30]. The lateral interface was characterized by high-resolution annular dark-field (ADF) scanning transmission electron microscopy (STEM). An atomically sharp and covalently bonded zigzag interface between the $WSe_2$ and $WS_{1.16}Se_{0.84}$ domains can be observed (Fig. 1a and a large-area TEM image in Supplementary Fig. 3), due to different scattering strength with electrons by W, S and Se atoms. The photoluminescence (PL) line scan (Fig. 1b) across the lateral interface showed that the position of PL peak for $WSe_2$ and $WS_{1.16}Se_{0.84}$ remained constant within each region, lending additional support for a sharp interface. The emission energies of the A excitons in $WSe_2$ and $WS_{1.16}Se_{0.84}$ domains were determined to be 1.64 and 1.81 eV, respectively (Fig. 1b and Supplementary Fig. 4).

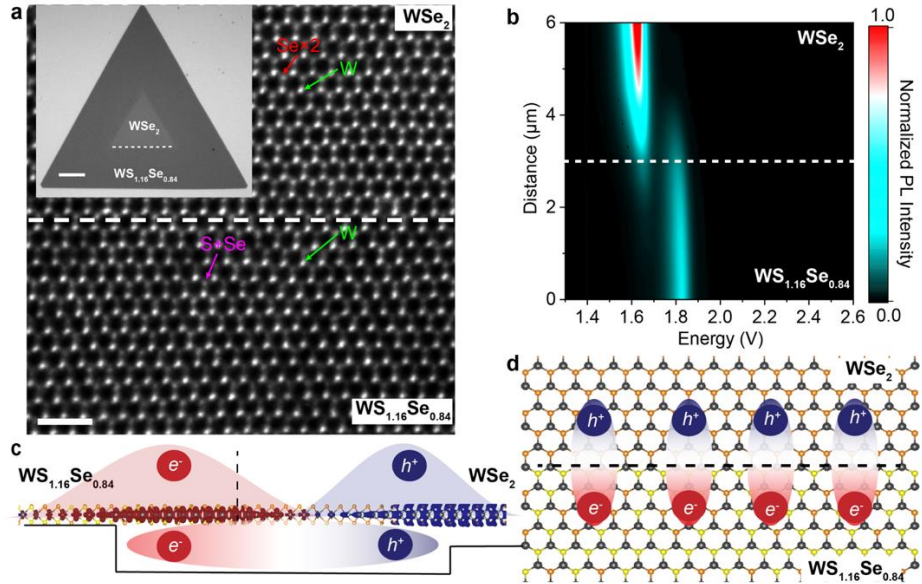

**Fig. 1| Interfacial excitons at a type II $WSe_2$-$WS_{1.16}Se_{0.84}$ lateral interface. a,** STEM image of a $WSe_2$-$WS_{1.16}Se_{0.84}$ lateral interface. The dashed line indicates the interface. Scale bar represents 1 nm. Inset: Optical micrograph of a typical lateral heterojunction with the interface marked by a dashed line. Scale bar represents 20 μm. **b,** PL line scan image across the same interface displayed in **a**. The white dashed line indicates the interface. **c,** Local density of state calculated using DFT integrated over of conduction band minimum (CBM) and valance band maximum (VBM) of ± 25 meV (corresponding to $k_BT$ at 290 K) of a $WSe_2$-$WS_{1.15}Se_{0.85}$ lateral interface model showing where the electrons and holes are localized. The isovalue of $5 \times 10^{-4}$ $e^-$ $Å^{-3}$ was selected. The interfacial exciton represents the ground state of the lateral heterostructure and the lateral interface can be viewed as an ultrathin 1D quantum wire. **d,** Schematic showing the alignment of the dipole moment of the interfacial excitons.

DFT calculations (see Methods for more details) was carried out to elucidate the electronic structure at the interface. Our calculations confirmed a type II band alignment (Supplementary Fig. 5) consistent with previous reports[19,31]. The band offset was estimated to be 0.2 eV from the DFT calculation (Supplementary Fig. 5). As shown in Fig. 1c, electrons (conduction band minimum, CBM) and holes (valance band maximum, VBM) were found to be spatially separated into WS$_{1.16}$Se$_{0.84}$ and WSe$_2$ parts of the interface with an e-h distance about 5.7 nm. The CT exciton is the ground state of the lateral heterostructure[22] and the interface can be viewed as an ultrathin excitonic quantum wire. Recent scanning tunneling microscopy (STM) measurements revealed ultra-narrow electronic transition region of only ~ 3 nm in width for WSe$_2$-WS$_2$ lateral heterostructures[32]. As the electron and hole wavefunctions are preferentially located at either side of the interface, both the transition and permanent dipole moment of interfacial CT excitons is expected to be perpendicular to the interface (Fig. 1d).

**1D Interfacial Exciton Dynamics**. We performed spatially-resolved and polarization-dependent pump-probe transient reflection (TR) spectroscopy with a time resolution of ~ 200 fs (Methods, Supplementary Fig. 6 and Fig. 7) to probe the dynamics of the interfacial CT excitons. Fig. 2a compares the TR spectra (plotted in pump-induced change of reflectance, $\Delta R/R$) at the interface and in the WS$_{1.16}$Se$_{0.84}$ and WSe$_2$ regions. The spatial resolution of the TR spectroscopy measurements is about 200 nm (Supplementary Fig. 8). The pump photon energy was above the bandgap at 3.14 eV. All TR spectra at zero delay time exhibited a derivative-shape near the A exciton resonance, due to a combination of Pauli blocking and transient bandgap renormalization resulting from the screening effect from photo-generated carriers or excitons[5,33,34]. Photoinduced bleaching bands at 1.81 and 1.64 eV were observed for WS$_{1.16}$Se$_{0.84}$ and WSe$_2$, respectively, agreeing well with the optical bandgaps determined from PL measurements (Fig. 1b). Notably, when the pump and probe beams were focused at the interface, a TR spectrum, distinct from that of WS$_{1.16}$Se$_{0.84}$ and WSe$_2$ regions, was observed (Fig. 2a), with a bleaching band ~ 30 meV redshifted from that of WS$_{1.16}$Se$_{0.84}$. The actual shift at the interface could be much larger as the pump and probe beam sizes were much wider than the interface; thus, the measured spectrum was a convolution of the interface and the surrounding area. One possible explanation for the redshift is the reduction of the bandgap of $WS_{1.16}Se_{0.84}$ at the interface by strain[32,35] induced by the lattice mismatch (~ 2%) between $WS_{1.16}Se_{0.84}$ and $WSe_2$ as confirmed by the selected-area electron diffraction measurement (Supplementary Fig. 9)[36]. A band gap reduction up of 200 meV due to 3.2% uniaxial strain was observed by STM measurements along the interface in a $WS_2$-$WSe_2$ lateral heterostructure.[32] The strain was estimated to be ~ 0.8 % at the $WS_{1.16}Se_{0.84}$ -$WSe_2$ interface (Fig. 2b). Stark effect due to a built-in electric field at the interface could also lead to a redshift of exciton energy[37]. We estimate the Stark effect by calculating the built-in electric field strength from the surface potential mapping using Kelvin force probe microscopy (KFPM) (Supplementary Note 1 and Supplementary Fig. 10). The maximum built-in electric field is determined to be $7 \times 10^4$ V/m, which gives a Stark energy shift of ~ $10^{-4}$ meV[37], much smaller than the 30 meV shift observed. Thus, we conclude that the redshift is mostly likely due to strain. We ascribed the photoinduced bleach band at 1.81 eV to the electrons residing in the interfacial $WS_{1.16}Se_{0.84}$ region after the formation of CT excitons.

Based on the spectra shown in Fig. 2a, we employed a probe energy of 1.81 eV for pump-probe TR microscopy to visualize the interface. At 1.81 eV, the $WS_{1.16}Se_{0.84}$ region exhibited a relatively weak positive TR signal while the lateral interface has a strong negative TR signal. There was no detectable TR signal in the $WSe_2$ region, because the probe photon energy was far away from the A exciton resonance of $WSe_2$. The reduced background signal from the surrounding $WS_{1.16}Se_{0.84}$ and $WSe_2$ areas at a probe energy of 1.81 eV allowed the lateral interface to be visualized by the pump-probe microscopy (Fig. 2c and Supplementary Fig. 11), despite the pump and probe beam size (~ 200 nm) being much larger than the interface. In these measurements, the pump and probe beams were overlapped in space and the sample was scanned with a piezo scanner. The image taken at zero delay time is shown in Fig. 2c, where the 1D interface can be clearly seen and indicated by the dashed line. Significantly slower carrier recombination was observed at the interface, as shown in Fig. 2d; a recombination time of 253 ± 41 ps at the interface compared with that of (45 ± 7 ps) for $WSe_2$ and (30 ± 2 ps) for $WS_{1.16}Se_{0.84}$. The slower recombination supports the spatial indirect nature of interfacial CT excitons. Strains at the interface can reduce intervalley scattering, which could also contribute to the slower recombination[35].

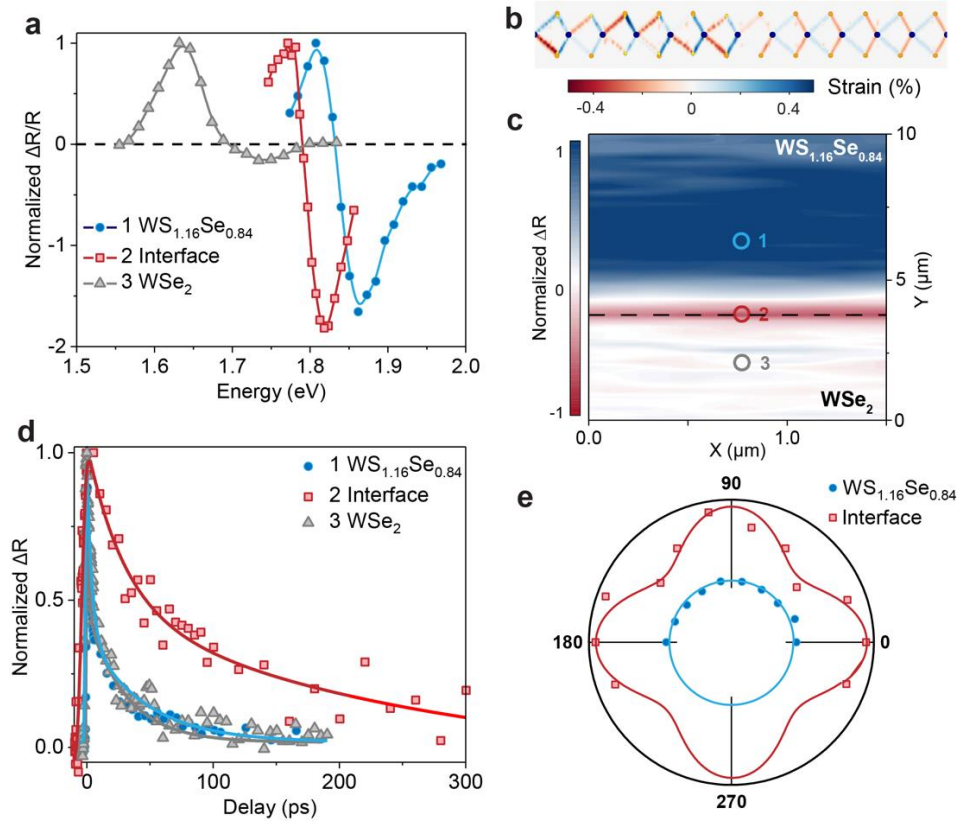

**Fig. 2| 1D interfacial exciton dynamics. a**, TR spectrum measured in the $WSe_2$, $WS_{1.16}Se_{0.84}$ and the interface regions of the lateral heterostructure. **b,** Calculated strain distribution at the interface. **c,** Pump-probe morphology image of the lateral heterojunction at 0 ps with spatially overlapped pump and probe beams. The interface can be visualized by a probe photon energy of 1.81 eV, because of the redshifted TR spectrum at the interfaces as shown in **a**. The exciton density was $2.0 \times 10^{12}$ cm$^{-2}$. **d,** TR dynamics measured at different positions (1, 2 and 3) marked in **c**. Solid lines are fits using a bi-exponential decay function, convoluted with a Gaussian response function. **e,** Angular-dependent TR polarization measured at positions 1 and 2 taken with probe linear polarized along the interface while rotating the pump polarization using a half-wave plate. Solid lines are fits using $I = a \times cos^2(\theta - \theta_0) + b$. The pump photon energy was 3.14 eV for all measurements.

The formation of the 1D interfacial excitons was further confirmed by polarization dependent TR measurements, where the probe was linearly polarized along the interface and the polarization of the pump was rotated. Fig. 2e shows the polar plot of $\Delta R/R$ as function of azimuth angle ($\theta$) measured at the interface. At the interface, $\Delta R/R$ displayed a 4-fold symmetry with two lobes ($\theta = 90°$ or $270°$ and $\theta = 0°$ or $180°$). The fact that we observed maxima at $\theta = 90°$ and $270°$ provides strong support for the interfacial CT excitons schematically shown in Fig. 1d, whose transition dipoles are aligned perpendicularly to the interface due to the spatially separated electron and hole wavefunctions across the interface. Without the formation of CT exciton, only a 2-fold symmetry would be expected in the polarization dependence with maxima at $\theta = 0°$ and $180°$ for strain-induced electronic state due to the uniaxial strain along the interface. In contrast, $\Delta R/R$ in the $WS_{1.16}Se_{0.84}$ region exhibited an isotropic pattern with $\theta$ as expected from exciton dipoles randomly orientated in the 2D plane. The polarization was also consistent with theoretical predictions of interfacial excitons coupling to linearly polarized light instead of circularly polarized light[22].

**First-Order Exciton Mott Transition Revealed by Ultrafast Microscopy.** Next, we employed spatially offset pump-probe microscopy to image carrier and exciton transport by fixing the pump beam at the interface and scanning the probe beam relative to the pump in space (see Methods). The polarization of the pump and probe beams were set to be perpendicular to the interface. $\Delta R/R$ at 1.81 eV, which was attributed to the electrons residing in $WS_{1.16}Se_{0.84}$ at the interface after the formation of the interfacial excitons, was mapped as function of probe position. $\Delta R/R$ was linearly proportional to the injected e-h density $N_0$ (Supplementary Fig. 12) and the spatially dependent TR signal was used as a measure of the carrier/exciton distribution at the interface. The details on the determination of $N_0$ can be found in Supplementary Note 2 and Supplementary Fig. 13. Figure 3a shows the pump-probe images taken at the lateral interface at 0, 5, 10, and 20 ps time delays with different $N_0$, illustrating the transport of excitons and charge carriers. We note that there were some uncertainties in determining exciton density at the interface experimentally, as these densities did not include the excitons funneled to the interface and, thus, the values here represent a lower bound estimate. However, exciton diffusion in WSe$_2$ and WS$_{1.16}$Se$_{0.84}$ regions is slow, as discussed below, the funneled population should be negligible initially, but becomes more important at longer delay times. As shown in Supplementary Fig. 14, about 20% of the population generated 20 nm away from the interface can be funneled to the interface within 8 ps.

The transport is highly anisotropic along the interface, consistent with the interfacial excitons being confined in a 1D channel. Most interestingly, as $N_0$ increased above $4.2 \times 10^{12}$ cm$^{-2}$, a very rapid expansion of the carrier population was observed (Fig. 3a). In direct contrast, exciton transport in the WS$_{1.16}$Se$_{0.84}$ and WSe$_2$ regions away from the interface was isotropic with negligible $N_0$ dependence (Fig. 3b and Supplementary Fig. 15). Exciton motion in the WS$_{1.16}$Se$_{0.84}$ and WSe$_2$ regions was limited even at $N_0$ as high as $1.2 \times 10^{13}$ cm$^{-2}$ (Fig. 3b and Supplementary Fig. 16), due to the combination of relatively short life time and small diffusion constant in the CVD grown materials (~ 0.1 cm$^2$ s$^{-1}$)[38].

To obtain a more quantitative understanding, the pump-probe microscopy images in Figs. 3a and 3b were modeled numerically. At zero delay time, exciton population created by a Gaussian pump beam can be described by a spatial distribution of $N(x, 0) = N_0 exp\left[-\frac{(x-x_0)^2}{2\sigma_{x,0}^2} - \frac{(y-y_0)^2}{2\sigma_{y,0}^2}\right]$ and the population profile broadens to $N_t exp[-\frac{(x-x_0)^2}{2\sigma_{x,t}^2} - \frac{(y-y_0)^2}{2\sigma_{y,t}^2}]$ at a later time $t$ (Fig. 3c-e). The mean distance $L$ travelled by excitons or charge carriers is given by $\sqrt{\sigma_t^2 - \sigma_0^2}$. The precision in determining exciton transport distance $L$ is not governed by the diffraction limit but by the smallest measurable change in the population profiles[39,40] (Supplementary Note 3). $\sigma_t^2 - \sigma_0^2$ along the interface at different $N_0$ was depicted in Fig. 4a. If transport was purely diffusive, $\sigma_t^2 - \sigma_0^2$ would grow linearly as function of delay time[41], with $N_0$- independent diffusion constant along the $x$ ($y$) direction given by $D_{x(y)} = \frac{\sigma_{x(y),t}^2 - \sigma_{x(y),0}^2}{2t}$. However, interfacial exciton migration accelerated as $N_0$ increased (Fig. 4a), deviating from a pure diffusive picture.

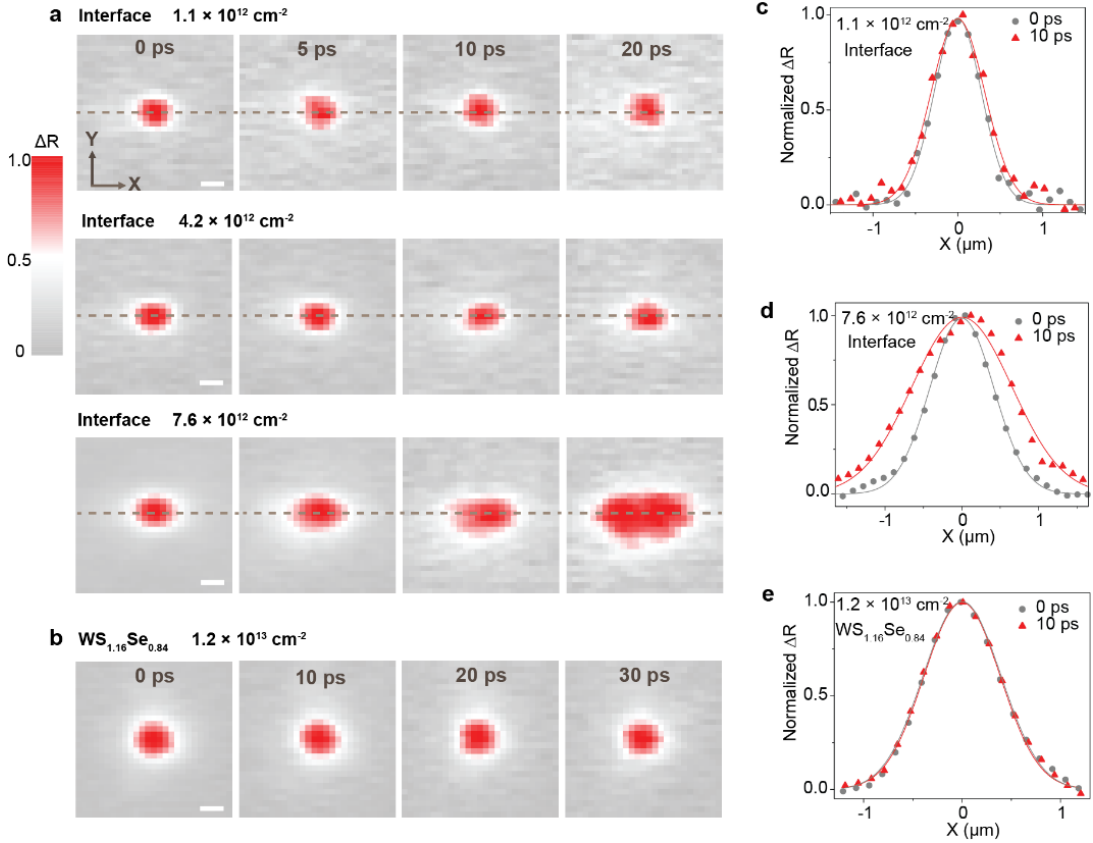

**Fig. 3| Many-body interactions driving transport in 1D. a,** Time-dependent pump-probe transport images taken at the lateral interface with different injected exciton densities ($N_0 = 1.1 \times 10^{12}$, $4.2 \times 10^{12}$, $7.6 \times 10^{12}$ cm$^{-2}$). The interface was marked by the dashed lines. More rapid carrier expansion was observed at $N_0 = 7.6 \times 10^{12}$ cm$^{-2}$. **b,** Time-dependent pump-probe transport images taken in the WS$_{1.16}$Se$_{0.84}$ region away from the interface. ($N_0 = 1.2 \times 10^{13}$ cm$^{-2}$). **c, d,** Carrier population profiles along the interface at different pump–probe delay times with injected exciton densities of $1.1 \times 10^{12}$ and $7.6 \times 10^{12}$ cm$^{-2}$, respectively. **e,** Carrier population profile of the WS$_{1.16}$Se$_{0.84}$ away from the interface with an injected exciton density of $1.2 \times 10^{13}$ cm$^{-2}$. Solid lines are fits using Gaussian functions. The pump and probe photon energies were 3.14 and 1.81 eV, respectively. Scale bars in **a** and **b** represent 0.5 μm. Solid lines in **d** and **e** are fits using Gaussian functions.

The strong dependence on $N_0$ suggests the key role of many-body effects in interfacial exciton transport. Remarkably, an abrupt increase in the time-dependence $\sigma_t^2 - \sigma_0^2$ can be seen at $N_0 \sim 5 \times 10^{12}$ cm$^{-2}$ (this abrupt increase was also observed in multiple lateral heterostructures as shown in Supplementary Fig.17). Above this density, some transport across the interfaces also can be observed (Supplementary Fig. 18). To better illustrate this transition, the effective diffusion constant at time zero $D_i = \frac{\partial[\sigma_t^2 - \sigma_0^2]}{2\partial t}$ ($t = 0$) along the interface was plotted as function of $N_0$ in Fig. 4c, showing two discontinuous regimes of transport. In contrast, no obvious density dependence was observed in the WS$_{1.16}$Se$_{0.84}$ region away from the interface (Fig. 4b) and $D_i$ remained constant over entire $N_0$ range (Fig. 4c).

The discontinuity in diffusivity suggests a first-order phase transition, which we attributed to a Mott transition from the insulating excitonic gas to a dense e-h plasma[6,42,43]. Increase of diffusivity has been previously reported as an indication of Mott transition in coupled quantum wells[43], TMDC monolayers[7] and heterostructures[6]. The Mott transition takes place when the exciton binding energy is reduced to zero, due to the strong screening of Columbic interaction when carrier density is increased beyond the critical density $N_{Mott}$ as schematic shown in Fig. 4e. $N_{Mott}$ is estimated by calculating the degree of ionization (α) of excitons using the Saha equation: $\frac{\alpha^2}{1-\alpha} = \frac{1}{n\lambda_T^2} exp\left(-\frac{\Delta E_{gap} - E_b}{k_B T}\right)$, where $\lambda_T = \frac{h}{\sqrt{2\pi m_r k_B T}}$ is the thermal de-Broglie length defined by electro-hole reduced mass ($m_r = \frac{m_e m_h}{m_e + m_h}$), $\Delta E_{gap}$ is the band gap renormalization with increased excitation density, $E_b$ is the exciton binding energy[11]. We used $\Delta E_{gap}$ and $E_b$ values from Ref.[11] and Ref.[22] and obtained α as function of exciton density, as shown in Fig. 4d. α displays an abrupt increase from 0.04 to 1, when the excitation density is increased from $5.0 \times 10^{12}$ to $6.0 \times 10^{12}$ cm$^{-2}$, in good agreement with transition revealed by carrier diffusivity measurements (Fig. 4c).

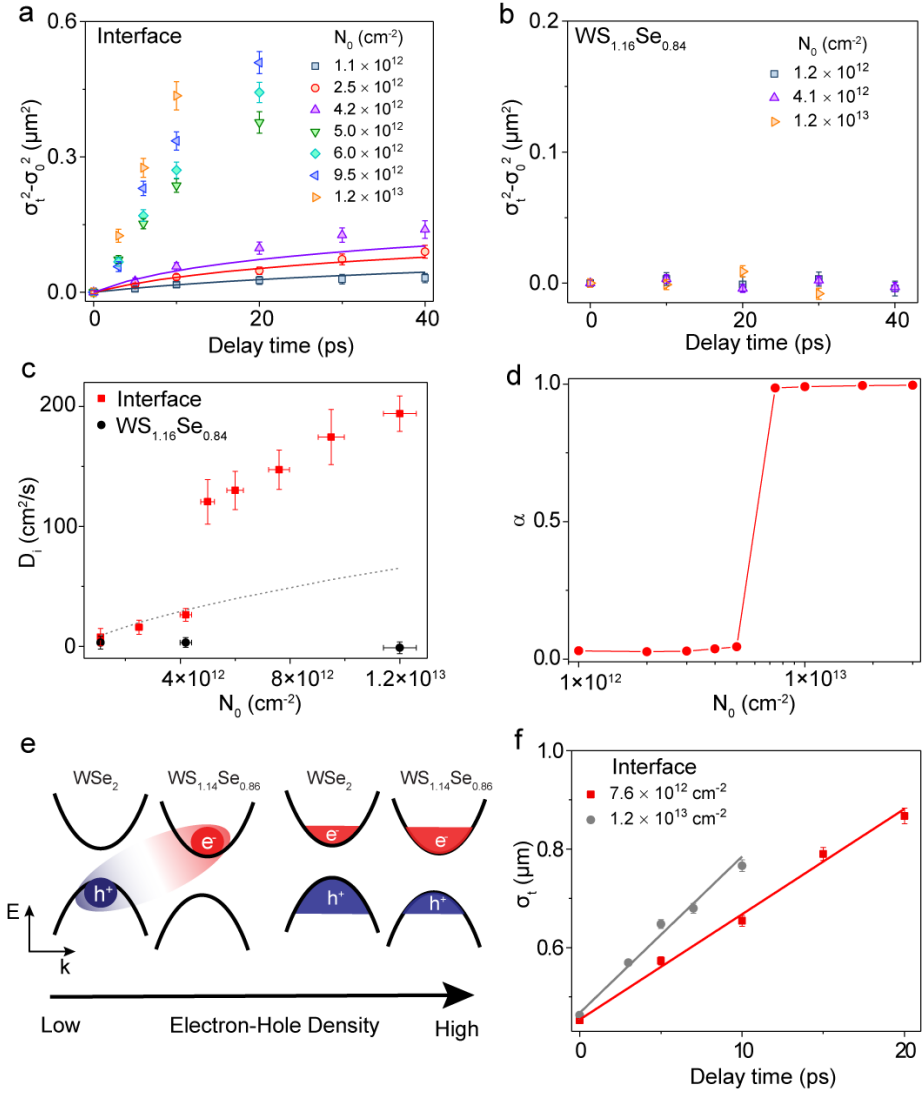

**Fig. 4| Exciton Mott transition at the lateral interface. a, b,** Exciton-density-dependent $\sigma_t^2 - \sigma_0^2$ as function of delay time along the interface and in the $WS_{1.16}Se_{0.84}$ region, respectively. Solid lines are simulation results from the drift-diffusion model of exciton transport described in the main text. **c,** Effective diffusion coefficient at time zero ($D_i$) as function of the injected exciton density ($N_0$) in the lateral interface and the $WS_{1.16}Se_{0.84}$ region. The dashed line indicates the simulated results from the drift-diffusion model of excitons as described in the main text, which fails to capture the abrupt increase in $D_i$ at $N_0$ above $5 \times 10^{12}$ cm$^{-2}$. **d,** Calculated degree of ionization ($\alpha$) as function of excitation density. **e,** Schematic of exciton Mott transition. **f,** Determination of drift velocity in the e-h plasma regime. Solid lines are linear fits.

It is possible that exciton-exciton annihilation or repulsion could also lead to apparent increase in exciton diffusion at high carrier density[44]; however, annihilation or repulsion should lead to a gradual change rather than the abrupt transition observed here, as shall be discussed further below. We also concluded that structural changes due to heating were minimal in these measurements by estimating the increase of lattice temperature by laser heating to be only ~ 16 K at the highest pump intensity (Supplementary Note 4). Interestingly, the Mott transition was not observed in the WSe$_2$ and WS$_{1.16}$Se$_{0.84}$ regions away from the interface under the same $N_0$. This difference is likely due to the much shorter intralayer exciton lifetime, which limits the achievable carrier density.

**Modeling of the Transport of Exciton Gas and E-H Plasma.** First, we discuss the transport of exciton gas below $N_{Mott}$. As illustrated in Fig. 1d, a key feature of the interfacial excitons are the permanent in-plane electric dipole moments aligned perpendicular to the interface. The aligned dipole moments result in repulsive interactions between the interfacial excitons, similar to the interactions between the dipolar excitons in TMDC vertical heterostructures[29,45] and coupled quantum wells[46]. At low exciton density, the interfacial potential fluctuations caused by defects, impurities, or alloying can limit the diffusion of CT excitons. As density increases, but still below the Mott transition, the repulsive interactions enhance exciton transport by effectively screening the energetic disorder, which was observed in our experiments as shown in Fig. 4a[29,46]. These repulsive interactions led to substantially faster transport of the interfacial excitons than the intralayer excitons in WS$_{1.16}$Se$_{0.84}$ region.

The exciton-density-dependent transport below $N_{Mott}$ is analogous to the exciton motion in a vertical vdW WS$_2$-WSe$_2$ heterostructure as reported in our previous work[29]. Thus, we employed a similar drift-diffusion model that includes exciton interactions and annihilation[29,47] (more details in the Supplementary Note 5) to simulate the interfacial exciton transport below the Mott transition. There are two contributions: one is from normal diffusion due to population gradient and the other is the drift motion resulting from exciton-exciton repulsion. The rate equation that describes the spatial and temporal distribution of excitons is given by: $\frac{\partial n(x,t)}{\partial t} = -\nabla \cdot$ $(J_{diff} + J_{drift}) - \frac{n(x,t)}{\tau} - \gamma n(x,t)^2$. Here, $\tau$ is the exciton lifetime and $\gamma$ is the annihilation rate. $J_{diff} = -D\nabla n(x,t)$ describes exciton diffusion and $J_{drift} = -n(x,t)\mu u_0 \nabla n(x,t)$ denotes the exciton drift flux due to the repulsive exciton-exciton interactions, where $\mu$ is the exciton mobility and $u_0$ is the exciton interaction energy and has contribution from both dipole-dipole interaction and exchange interaction[29]. As shown in Figs 4a and 4c, this model successfully captured the density- and time-dependent exciton transport for exciton density in the range of $1.1 \times 10^{12} - 4.2 \times 10^{12}$ cm$^{-2}$ using $D = 0.74$ cm$^2$s$^{-1}$, $u_0 = 3.9 \times 10^{-14}$ eVcm$^{-2}$ and $\gamma = 0.03$ cm$^2$s$^{-1}$. However, at $N_0$ above $5 \times 10^{12}$ cm$^{-2}$, this model fails to simulate the abrupt increase in carrier diffusivity (Fig. 4c and Supplementary Figs. 19 and 20), which is consistent with the fact that excitons do not exist above $N_{Mott}$ and the exciton-exciton interaction term is no longer valid.

The driving force for the expansion of the e-h plasma is known as Fermi pressure[48-50], i.e., at high carrier density above the onset of degeneracy, the quasi Fermi level of the electron and hole is strongly dependent on their density due to Pauli's exclusion principle. Because the carrier density was higher at the center than at the edge of the excitation spot, there was a gradient of chemical potential $\Phi$ in space, which acts as an effective electric field $\frac{\partial \Phi}{\partial x}$ that drives the carriers away from the center (Supplementary Note 6). Approximating a chemical potential as $\frac{ne^2d}{\varepsilon}$ (where $n$ is the carrier density, $\varepsilon$ is the dielectric constant and $d$ the e-h distance)[43], the maximum electric field induced by the chemical potential gradient is estimated to be 15 KV/cm at $n = 1.2 \times 10^{13}$ cm$^{-2}$ (9.4 KV/cm at $n = 7.6 \times 10^{12}$ cm$^{-2}$, Supplementary Fig. 21). The drift velocity of charge carriers can be extracted from $\sigma_t = v_d t$, where $v_d$ is the drift velocity, as shown in Fig.4f. $v_d$ of $3.2 \times 10^6$ cm s$^{-1}$ and $2.1 \times 10^6$ cm s$^{-1}$ can be obtained for a carrier density of $1.2 \times 10^{13}$ cm$^{-2}$ and $7.6 \times 10^{12}$ cm$^{-2}$, respectively. These velocities agreed well with the carrier drift velocities of WS$_2$ at high field calculated from Monte Carlo simulations (~ $3.5 \times 10^6$ cm s$^{-1}$ at 15 KV/cm)[51,52] with similar values for electrons and holes. $v_d$ is orders of magnitude larger than the sound velocity, which also rules out phonon winds being the driving force for e-h plasma expansion[53].

**Discussion**

The results presented here establish a first-order Mott transition of the 1D interfacial excitons in lateral TMDC heterostructures. Importantly, the involvement of a highly non-equilibrium e-h plasma phase represents an important distinction between the Mott transition observed here and the more conventional thermodynamic phase transition. The e-h plasma is not a thermodynamically stable phase, but rather its rapid expansion leads to fast reduction of e-h density and conversion back to the exciton phase. Therefore, the non-equilibrium e-h plasma phase can escape conventional steady-state measurements that integrate over long time, leading to an apparent continuous second-order transitions. The combined high spatial and temporal resolutions hold the keys in elucidating the first-order nature of such a non-equilibrium phase transition. Enhanced diffusivity above the Mott density has been reported in monolayer $MoS_2$ by Yu et al.[7] and in $MoSe_2$-$WSe_2$ by Wang et al.[6], both studied using PL microscopy. However, in these previous reports[6,7], lower diffusivities (~ 20 cm$^2$s$^{-1}$ at an injected exciton density of $10^{14}$ cm$^{-2}$ in Ref.[7]) were reported and no abrupt change was observed, likely due to the much lower time resolution (~ 100 ps in Ref.[6] and no time resolution in Ref.[7]). In contrast, the high time resolution ~ 200 fs used in our work was able to reveal an abrupt increase in effective diffusivity by almost an order of magnitude, approaching 200 cm$^2$ s$^{-1}$.

Thus, the ultrafast expansion of e-h plasma provides a plausible answer to the debate regarding whether the exciton Mott transition is a first-order (abrupt) or second-order (continuous) phase transition[2,3,24,42,54], because a first-order non-equilibrium transition can be mistaken as a second-order transition in steady-state experiments. As Mott originally proposed, the insulator to metal transition should be first-order at sufficiently low temperature[1]. A large body of theoretical works have verified the existence of first-order exciton Mott transitions below a critical temperature $T_c$ that is related to exciton binding energy in 1D, 2D, and bulk semiconductors[2,24,54]. The large exciton binding energy of TMDCs and their heterostructures, on the order of hundreds of meV, should in principle allow for the observation of first-order exciton Mott transition at room temperature[11]. However, there are conflicting experimental observations regarding the nature of Mott transition in TMDC materials. In fact, the only first-order transition so far was reported by Bataller et al.[8] showing an abrupt increase of PL intensity at Mott transition in suspended $MoS_2$; this observation was made possible by the elongation of carrier lifetime to hundreds of ns in the e-h plasma phase and a sample geometry that trapped e-h pairs to a certain region, which can establish a quasi-equilibrium condition. The lack of Mott transition in $WSe_2$ and $WS_{1.16}Se_{0.84}$ regions away from the interface underscores the importance of the long interfacial exciton lifetime in the realization of Mott transition. We hypothesized that another key factor could be the energy funneling to the interface. Because the interfacial exciton state is effectively an energy sink, exciton population can be replenished from the nearby $WSe_2$ and $WS_{1.16}Se_{0.84}$ regions at later delay time, as illustrated in Supplementary Fig. 14.

In conclusion, these results point to an exciting direction in designing TMDC lateral heterojunctions as "highways" for harnessing collective electronic phases. In particular, the ability to realize dense e-h plasma with extremely high mobility at room temperature is promising for high-speed electronic applications. Recent breakthroughs in materials synthesis have realized ordered arrays of lateral heterojunctions with precise composition and strain control[20], which could provide an entirely new materials platform for electronic devices based on e-h plasma. Finally, the energy funneling effects at these interfaces can allow for exciting opportunities in modulating the balance between driving and dissipation to realize a driven-dissipative phase transition in solid state systems[55].

## Methods

**Sample fabrication.** The WSe$_2$-WS$_{2(1-x)}$Se$_{2x}$ lateral heterojunctions were synthesized via a modified two-step chemical vapor deposition (CVD) method schematically shown in Supplementary Fig. 1. Briefly, a quartz boat with a 50:50 mixture of tungsten disulfide and tungsten diselenide powder (99.8%, Alfa Aesar) and a Si wafer with a 285-nm-thick oxide (1 cm × 3 cm) were placed at the center and the downstream of the furnace for the growth of the WS$_{2(1-x)}$Se$_{2x}$ monolayers respectively. The system was firstly cleaned by the high pure Ar gas (400 SCCM) for 15 min. The furnace was then heated to 1050 °C for the growth of WS$_{2(1-x)}$Se$_{2x}$ monolayers (10 mins). The as-prepared WS$_{2(1-x)}$Se$_{2x}$ monolayers were used as the new substrate for the lateral growth of the WSe$_2$ monolayers. The tungsten diselenide powder (99.8%, Alfa Aesar) and the substrate with the WS$_{2(1-x)}$Se$_{2x}$ monolayers were placed at the center and downstrem of the quartz tube respectively. A high pure flow of Ar gas was used as the carrier gas and the growth temperature was set at 1000°C. After the growth, the furnace was cooled down to the room temperature.

**Scanning transmission electron microscopy (STEM).** CVD grown WSe$_2$-WS$_{1.16}$Se$_{0.84}$ lateral heterojunctions were transferred onto a copper grid using the poly(methyl methacrylate) PMMA-assisted transfer method. The STEM measurements were carried out on JEOL ARM200F microscope operated at 200 kV and equipped with a probe-forming aberration corrector. For HAADF-STEM images, the inner and outer collection angles of the ADF detector are 68 and 280 mrad, respectively. The convergence semiangle is about 28 mard.

**Pump-probe reflection spectroscopy and microscopy.** All measurements were carried out at room temperature. Transient reflectance dynamics and pump-probe imaging measurements were taken using a home-built system, schematically shown in Supplementary Fig.6. Briefly, a Ti: Sapphire oscillator (Coherent Mira 900) pumped by a Verdi diode laser (Verdi V18) was used as the fundamental light source (1.57eV, 76 MHz, 200 fs). 70% of the pulse energy was fed into the optical parametric oscillator (OPO) (Coherent Mira OPO) to generate the probe beam at 1.81 eV, whereas the remaining 30% was doubled to 3.14 eV using a β-barium borate (BBO) crystal. The pump beam was modulated at 1 MHz using an acoustic optical modulator (AOM) (Model R21080-1DM, Gooch&Housego). Both pump and probe beams were spatially filtered. A 60X (NA = 0.95) objective was used to focus both pump and probe beams onto the sample, and the reflection light was then collected by the same objective and detected by an avalanche Si photodiode (Hamamatsu). The change in the probe reflection ($\Delta R$) induced by the pump was detected by a lock-in amplifier (HF2LI, Zurich Instrument). For TR dynamics measurements, the pump and probe beams were spatially overlapped and a mechanical translation stage (Thorlabs, LTS300) was used to delay the probe with respect to the pump. TR spectrum was measured by manually tuning the probe wavelength using the OPO. TR signal was expressed as $\Delta R / R$, where $\Delta R$ and $R$ are the change in the probe reflection induced by the pump and the probe reflection from the area without the sample respectively. For morphological imaging (Fig. 2c), pump and probe beams were spatially overlapped and a piezo-electric stage (P-527.3Cl, Physik Instrumente) was used to scan the sample with a step size of 50 nm. For carrier transport imaging (Fig.3), a galvanometer scanner (Thorlabs GVS012) was used to scan the probe beam relative to the pump beam in space to obtain the carrier population profiles.

**Density functional theory simulations.** To model the junction, we built a rectangular periodic system of WSe$_2$ monolayer with 2 unit cells in x direction and 40 unit cells in y direction, resulting in 240 atoms. Half of the model in y direction was kept as WSe$_2$ and the other half was randomly substituted with S atoms, to create a heterojunction of about 11 nm long and stoichiometry of WSe$_2$-WS$_{1.15}$Se$_{0.85}$ (close to the experimental stoichiometry). This model was relaxed such that the forces on lattice vectors and atomic positions were minimized, except the z coordinate of W atoms, to avoid rippling (experimental system is placed on WSe$_2$ layer, thus, rippling is not present). We used FHI-aims[56] employing the PBE functional[57] on tight tier 1 numeric atom-centered orbitals, including the non-self-consistent Tkatchenko-Scheffler correction for dispersion interactions[58], and scalar relativistic corrections (ZORA) on a 12 × 1 × 1 Γ-centered k-grid. The forces and stresses were minimized till below 0.01 eV per Å. The electronic band structure, the Mulliken-projections, and the density of states were calculated including spin-orbit coupling (SOC) and considering the dipole correction on a 12 × 1 × 1 Γ-centered k-grid. The atom-projected density of states (pDOS) was calculated using the tetrahedron method in FHI-aims. The pDOS has been convoluted with a Gaussian function (using broadening of 50 meV and keeping the total number of states constant). The grid of the pDOS values of W atoms obtained in this way has been interpolated using a cubic spline, as implemented in the scipy package[59]. The integrated local DOS (ILDOS) was calculated for different voltage values, corresponding to the energies from the Fermi level to the band edges ± 25 meV (corresponding to k$_B$T at 290 K). The strain distribution in Fig. 2b was calculated on the relaxed geometry of the heterostructure and comparing different bond lengths to those of the corresponding relaxed single WSe$_2$ and WS$_2$ monolayer unit cells.

## Acknowledgements

The optical spectroscopy and microscopy work at Purdue are supported by the US Department of Energy, Office of Basic Energy Sciences, through award DE-SC0016356. L.Y. also acknowledges support from the Purdue University Bilsland Dissertation Fellowship. L.Y. thanks Y. W and Z. G for their assistance in the instrument development. B.Z. and A.P. acknowledge the National Natural Science Foundation of China (Nos. 62090035 and U19A2090), the Key Program of Science and Technology Department of Hunan Province (2019XK2001, 2020XK2001), the China Postdoctoral Science Foundation (Nos. 2020M680112 and BX2021094), and the Science and Technology Innovation Program of Hunan Province (No. 2020RC2028). R.K, T.B, and A.B.K. gratefully acknowledge the Gauss Centre for Supercomputing e.V. (www.gauss-centre.eu) for funding this project by providing computing time through the John von Neumann Institute for Computing (NIC) on the GCS Supercomputer JUWELS at Jülich Supercomputing Centre (JSC) and the ZIH Dresden supercomputing center. T.B. and A.B.K acknowledge financial support by Deutsche Forschungsgemeinschaft (DFG, German Research Foundation) within SFB1415 project number 417590517 and the association with the SPP2244 (2DMP).

## Author contributions

L.H., L.Y., and A.P. designed the experiments, L.Y. and Q.Z. carried out the optical measurements, B.Z., A.P. and C.M. grew the samples and performed the electron microscopy characterization, R.K., T.B., and A.B.K. carried out and analyzed the DFT calculations, L.Y., Q.Z., S.D., and L.H. analyzed experimental data, L.Y. and L.H. wrote the manuscript with input from all the authors.

## Competing interests

The authors declare no competing interests.

## References

1. Mott, N. F. Metal-Insulator Transition. *Reviews of Modern Physics* **40**, 677-683, (1968).

2. Haug, H. On the phase transitions for the electronic excitations in semiconductors. *Physik B Condensed Matter and Quanta* **24**, 351-360, (1976).
3. Koch, S. W., Hoyer, W., Kira, M. & Filinov, V. S. Exciton ionization in semiconductors. *physica status solidi (b)* **238**, 404-410, (2003).
4. Jeffries, C. D. Electron-Hole Condensation in Semiconductors. *Science* **189**, 955-964, (1975).
5. Chernikov, A., Ruppert, C., Hill, H. M., Rigosi, A. F. & Heinz, T. F. Population inversion and giant bandgap renormalization in atomically thin $WS_2$ layers. *Nature Photonics* **9**, 466-470, (2015).
6. Wang, J., Shi, Q. H., Shih, E. M., Zhou, L., Wu, W. J., Bai, Y. S., Rhodes, D., Barmak, K., Hone, J., Dean, C. R. & Zhu, X. Y. Diffusivity Reveals Three Distinct Phases of Interlayer Excitons in MoSe2/WSe2 Heterobilayers. *Physical Review Letters* **126**, 106804, (2021).
7. Yu, Y., Yu, Y., Li, G., Puretzky, A. A., Geohegan, D. B. & Cao, L. Giant enhancement of exciton diffusivity in two-dimensional semiconductors. *Science Advances* **6**, eabb4823, (2020).
8. Bataller, A. W., Younts, R. A., Rustagi, A., Yu, Y., Ardekani, H., Kemper, A., Cao, L. & Gundogdu, K. Dense Electron–Hole Plasma Formation and Ultralong Charge Lifetime in Monolayer $MoS_2$ via Material Tuning. *Nano Letters* **19**, 1104-1111, (2019).
9. Dendzik, M., Xian, R. P., Perfetto, E., Sangalli, D., Kutnyakhov, D., Dong, S., Beaulieu, S., Pincelli, T., Pressacco, F., Curcio, D., Agustsson, S. Y., Heber, M., Hauer, J., Wurth, W., Brenner, G., Acremann, Y., Hofmann, P., Wolf, M., Marini, A., Stefanucci, G., Rettig, L. & Ernstorfer, R. Observation of an Excitonic Mott Transition Through Ultrafast Core- *cum* -Conduction Photoemission Spectroscopy. *Physical Review Letters* **125**, 096401, (2020).
10. Kudlis, A. & Iorsh, I. Modeling excitonic Mott transitions in two-dimensional semiconductors. *Physical Review B* **103**, 115307, (2021).
11. Steinhoff, A., Florian, M., Rösner, M., Schönhoff, G., Wehling, T. O. & Jahnke, F. Exciton fission in monolayer transition metal dichalcogenide semiconductors. *Nature Communications* **8**, 1166, (2017).
12. Arp, T. B., Pleskot, D., Aji, V. & Gabor, N. M. Electron–hole liquid in a van der Waals heterostructure photocell at room temperature. *Nature Photonics* **13**, 245-250, (2019).
13. Yu, Y., Bataller, A. W., Younts, R., Yu, Y., Li, G., Puretzky, A. A., Geohegan, D. B., Gundogdu, K. & Cao, L. Room-Temperature Electron–Hole Liquid in Monolayer MoS2. *ACS Nano* **13**, 10351-10358, (2019).
14. Gong, Y., Lin, J., Wang, X., Shi, G., Lei, S., Lin, Z., Zou, X., Ye, G., Vajtai, R., Yakobson, B. I., Terrones, H., Terrones, M., Tay, Beng K., Lou, J., Pantelides, S. T., Liu, Z., Zhou, W. & Ajayan, P. M. Vertical and in-plane heterostructures from $WS_2$/$MoS_2$ monolayers. *Nature Materials* **13**, 1135-1142, (2014).
15. Huang, C., Wu, S., Sanchez, A. M., Peters, J. J. P., Beanland, R., Ross, J. S., Rivera, P., Yao, W., Cobden, D. H. & Xu, X. Lateral heterojunctions within monolayer $MoSe_2$–$WSe_2$ semiconductors. *Nature Materials* **13**, 1096-1101, (2014).
16. Duan, X., Wang, C., Shaw, J. C., Cheng, R., Chen, Y., Li, H., Wu, X., Tang, Y., Zhang, Q., Pan, A., Jiang, J., Yu, R., Huang, Y. & Duan, X. Lateral epitaxial growth of two-dimensional layered semiconductor heterojunctions. *Nature Nanotechnology* **9**, 1024-1030, (2014).
17. Li, M.-Y., Shi, Y., Cheng, C.-C., Lu, L.-S., Lin, Y.-C., Tang, H.-L., Tsai, M.-L., Chu, C.-W., Wei, K.-H., He, J.-H., Chang, W.-H., Suenaga, K. & Li, L.-J. Epitaxial growth of a monolayer WSe2-MoS2 lateral p-n junction with an atomically sharp interface. *Science* **349**, 524-528, (2015).
18. Zhang, Z., Chen, P., Duan, X., Zang, K., Luo, J. & Duan, X. Robust epitaxial growth of two-dimensional heterostructures, multiheterostructures, and superlattices. *Science* **357**, 788, (2017).
19. Zheng, B., Ma, C., Li, D., Lan, J., Zhang, Z., Sun, X., Zheng, W., Yang, T., Zhu, C., Ouyang, G., Xu, G., Zhu, X., Wang, X. & Pan, A. Band Alignment Engineering in Two-Dimensional Lateral Heterostructures. *Journal of the American Chemical Society* **140**, 11193-11197, (2018).
20. Xie, S., Tu, L., Han, Y., Huang, L., Kang, K., Lao, K. U., Poddar, P., Park, C., Muller, D. A., DiStasio, R. A. & Park, J. Coherent, atomically thin transition-metal dichalcogenide superlattices with engineered strain. *Science* **359**, 1131-1136, (2018).

21. Sahoo, P. K., Memaran, S., Xin, Y., Balicas, L. & Gutiérrez, H. R. One-pot growth of two-dimensional lateral heterostructures via sequential edge-epitaxy. *Nature* **553**, 63-67, (2018).
22. Lau, K. W., Calvin, Gong, Z., Yu, H. & Yao, W. Interface excitons at lateral heterojunctions in monolayer semiconductors. *Physical Review B* **98**, 115427, (2018).
23. Ávalos-Ovando, O., Mastrogiuseppe, D. & Ulloa, S. E. Lateral interfaces of transition metal dichalcogenides: A stable tunable one-dimensional physics platform. *Physical Review B* **99**, 035107, (2019).
24. Yoshioka, T. & Asano, K. Exciton-Mott Physics in a Quasi-One-Dimensional Electron-Hole System. *Physical Review Letters* **107**, 256403, (2011).
25. Murakami, Y. & Kono, J. Nonlinear Photoluminescence Excitation Spectroscopy of Carbon Nanotubes: Exploring the Upper Density Limit of One-Dimensional Excitons. *Physical Review Letters* **102**, 037401, (2009).
26. Yoshita, M., Hayamizu, Y., Akiyama, H., Pfeiffer, L. N. & West, K. W. Exciton-plasma crossover with electron-hole density in T-shaped quantum wires studied by the photoluminescence spectrograph method. *Physical Review B* **74**, 165332, (2006).
27. Guo, Z., Wan, Y., Yang, M., Snaider, J., Zhu, K. & Huang, L. Long-range hot-carrier transport in hybrid perovskites visualized by ultrafast microscopy. *Science* **356**, 59-62, (2017).
28. Zhu, T., Yuan, L., Zhao, Y., Zhou, M., Wan, Y., Mei, J. & Huang, L. Highly mobile charge-transfer excitons in two-dimensional $WS_2$/tetracene heterostructures. *Science Advances* **4**, eaao3104, (2018).
29. Yuan, L., Zheng, B., Kunstmann, J., Brumme, T., Kuc, A. B., Ma, C., Deng, S., Blach, D., Pan, A. & Huang, L. Twist-angle-dependent interlayer exciton diffusion in $WS_2$–$WSe_2$ heterobilayers. *Nature Materials* **19**, 617-623, (2020).
30. Cao, Z., Harb, M., Lardhi, S. & Cavallo, L. Impact of Interfacial Defects on the Properties of Monolayer Transition Metal Dichalcogenide Lateral Heterojunctions. *The Journal of Physical Chemistry Letters* **8**, 1664-1669, (2017).
31. Li, L., Zheng, W., Ma, C., Zhao, H., Jiang, F., Ouyang, Y., Zheng, B., Fu, X., Fan, P., Zheng, M., Li, Y., Xiao, Y., Cao, W., Jiang, Y., Zhu, X., Zhuang, X. & Pan, A. Wavelength-Tunable Interlayer Exciton Emission at the Near-Infrared Region in van der Waals Semiconductor Heterostructures. *Nano Letters* **20**, 3361-3368, (2020).
32. Herbig, C., Zhang, C., Mujid, F., Xie, S., Pedramrazi, Z., Park, J. & Crommie, M. F. Local Electronic Properties of Coherent Single-Layer $WS_2$/$WSe_2$ Lateral Heterostructures. *Nano Letters* **21**, 2363-2369, (2021).
33. Pogna, E. A. A., Marsili, M., De Fazio, D., Dal Conte, S., Manzoni, C., Sangalli, D., Yoon, D., Lombardo, A., Ferrari, A. C., Marini, A., Cerullo, G. & Prezzi, D. Photo-Induced Bandgap Renormalization Governs the Ultrafast Response of Single-Layer $MoS_2$. *ACS Nano* **10**, 1182-1188, (2016).
34. Yuan, L., Chung, T.-F., Kuc, A., Wan, Y., Xu, Y., Chen, Y. P., Heine, T. & Huang, L. Photocarrier generation from interlayer charge-transfer transitions in $WS_2$-graphene heterostructures. *Science Advances* **4**, e1700324, (2018).
35. Aslan, O. B., Deng, M. & Heinz, T. F. Strain tuning of excitons in monolayer $WSe_2$. *Physical Review B* **98**, 115308, (2018).
36. Zhang, C., Li, M.-Y., Tersoff, J., Han, Y., Su, Y., Li, L.-J., Muller, D. A. & Shih, C.-K. Strain distributions and their influence on electronic structures of $WSe_2$–$MoS_2$ laterally strained heterojunctions. *Nature Nanotechnology* **13**, 152-158, (2018).
37. Cavalcante, L. S. R., da Costa, D. R., Farias, G. A., Reichman, D. R. & Chaves, A. Stark shift of excitons and trions in two-dimensional materials. *Physical Review B* **98**, 245309, (2018).
38. Yuan, L., Wang, T., Zhu, T., Zhou, M. & Huang, L. Exciton Dynamics, Transport, and Annihilation in Atomically Thin Two-Dimensional Semiconductors. *The Journal of Physical Chemistry Letters* **8**, 3371-3379, (2017).

39. Akselrod, G. M., Deotare, P. B., Thompson, N. J., Lee, J., Tisdale, W. A., Baldo, M. A., Menon, V. M. & Bulović, V. Visualization of exciton transport in ordered and disordered molecular solids. *Nature Communications* **5**, 3646, (2014).
40. Sung, J., Schnedermann, C., Ni, L., Sadhanala, A., Chen, R. Y. S., Cho, C., Priest, L., Lim, J. M., Kim, H.-K., Monserrat, B., Kukura, P. & Rao, A. Long-range ballistic propagation of carriers in methylammonium lead iodide perovskite thin films. *Nature Physics* **16**, 171-176, (2020).
41. Zhu, T., Snaider, J. M., Yuan, L. & Huang, L. Ultrafast Dynamic Microscopy of Carrier and Exciton Transport. *Annual Review of Physical Chemistry* **70**, 219-244, (2019).
42. Stern, M., Garmider, V., Umansky, V. & Bar-Joseph, I. Mott Transition of Excitons in Coupled Quantum Wells. *Physical Review Letters* **100**, 256402, (2008).
43. Stern, M., Garmider, V., Segre, E., Rappaport, M., Umansky, V., Levinson, Y. & Bar-Joseph, I. Photoluminescence Ring Formation in Coupled Quantum Wells: Excitonic Versus Ambipolar Diffusion. *Physical Review Letters* **101**, 257402, (2008).
44. Kulig, M., Zipfel, J., Nagler, P., Blanter, S., Schüller, C., Korn, T., Paradiso, N., Glazov, M. M. & Chernikov, A. Exciton Diffusion and Halo Effects in Monolayer Semiconductors. *Physical Review Letters* **120**, 207401, (2018).
45. Li, W., Lu, X., Dubey, S., Devenica, L. & Srivastava, A. Dipolar interactions between localized interlayer excitons in van der Waals heterostructures. *Nature Materials* **19**, 624-629, (2020).
46. Vögele, X. P., Schuh, D., Wegscheider, W., Kotthaus, J. P. & Holleitner, A. W. Density Enhanced Diffusion of Dipolar Excitons within a One-Dimensional Channel. *Physical Review Letters* **103**, 126402, (2009).
47. Ivanov, A. L. Quantum diffusion of dipole-oriented indirect excitons in coupled quantum wells. *Europhysics Letters (EPL)* **59**, 586-591, (2002).
48. van Driel, H. M. Kinetics of high-density plasmas generated in Si by 1.06- and 0.53- $\mu m$ picosecond laser pulses. *Physical Review B* **35**, 8166-8176, (1987).
49. Majumder, F. A., Swoboda, H.-E., Kempf, K. & Klingshirn, C. Electron-hole plasma expansion in the direct-band-gap semiconductors CdS and CdSe. *Physical Review B* **32**, 2407-2418, (1985).
50. Ziebold, R., Witte, T., Hübner, M. & Ulbrich, R. G. Direct observation of Fermi-pressure-driven electron-hole plasma expansion in GaAs on a picosecond time scale. *Physical Review B* **61**, 16610-16618, (2000).
51. Jin, Z., Li, X., Mullen, J. T. & Kim, K. W. Intrinsic transport properties of electrons and holes in monolayer transition-metal dichalcogenides. *Physical Review B* **90**, 045422, (2014).
52. Smithe, K. K. H., English, C. D., Suryavanshi, S. V. & Pop, E. High-Field Transport and Velocity Saturation in Synthetic Monolayer MoS2. *Nano Letters* **18**, 4516-4522, (2018).
53. Steranka, F. M. & Wolfe, J. P. Phonon-Wind-Driven Electron-Hole Plasma in Si. *Physical Review Letters* **53**, 2181-2184, (1984).
54. Guerci, D., Capone, M. & Fabrizio, M. Exciton Mott transition revisited. *Physical Review Materials* **3**, 054605, (2019).
55. Tomita, T., Nakajima, S., Danshita, I., Takasu, Y. & Takahashi, Y. Observation of the Mott insulator to superfluid crossover of a driven-dissipative Bose-Hubbard system. *Science Advances* **3**, e1701513, (2017).
56. Blum, V., Gehrke, R., Hanke, F., Havu, P., Havu, V., Ren, X., Reuter, K. & Scheffler, M. Ab initio molecular simulations with numeric atom-centered orbitals. *Computer Physics Communications* **180**, 2175-2196, (2009).
57. Perdew, J. P., Burke, K. & Ernzerhof, M. Generalized Gradient Approximation Made Simple. *Physical Review Letters* **77**, 3865-3868, (1996).
58. Tkatchenko, A. & Scheffler, M. Accurate Molecular Van Der Waals Interactions from Ground-State Electron Density and Free-Atom Reference Data. *Physical Review Letters* **102**, 073005, (2009).
59. Virtanen, P., Gommers, R., Oliphant, T. E., Haberland, M., Reddy, T., Cournapeau, D., Burovski, E., Peterson, P., Weckesser, W., Bright, J., van der Walt, S. J., Brett, M., Wilson, J., Millman, K. J., Mayorov, N., Nelson, A. R. J., Jones, E., Kern, R., Larson, E., Carey, C. J., Polat, İ., Feng, Y., Moore, E.

W., VanderPlas, J., Laxalde, D., Perktold, J., Cimrman, R., Henriksen, I., Quintero, E. A., Harris, C. R., Archibald, A. M., Ribeiro, A. H., Pedregosa, F. & van Mulbregt, P. SciPy 1.0: fundamental algorithms for scientific computing in Python. *Nature Methods* **17**, 261-272, (2020).